\documentclass{article}
\usepackage[utf8]{inputenc}
\usepackage{graphicx}
\usepackage{acronym}
\usepackage{wrapfig}
\usepackage{lscape}
\usepackage{rotating}

\title{COVID-19 Outbreak in Pakistan: Model-Driven Impact Analysis and Guidelines}
\author{Kashif Zia, Faculty of Computing and IT, Sohar University, Oman
\\
Umar Farooq, University of Science and Technology Bannu, Pakistan
}
\date{March 31, 2020}

\acrodef{covid19}[COVID-19]{Coronavirus disease 2019}
\acrodef{sars}[SARS]{Severe Acute Respiratory Syndrome}

\begin{document}

\maketitle

\begin{abstract}
Motivated by the rapid spread of COVID-19 all across the globe, we have performed simulations of a system dynamic epidemic spread model in different possible situations. The simulation, not only captures the model dynamic of the spread of the virus, but also, takes care of population and mobility data. The model is calibrated based on epidemic data and events specifically of Pakistan, which can easily be generalized. The simulation results are quite disturbing, indicating that, even with stringent social distancing and testing strategies and for a quite long time (even beyond one year), the spread would be significant (in tens of thousands). The real alarm is when some of these measures got leaked for a short time within this duration, which may result in catastrophic situation when millions of people would be infected.  

\end{abstract}

\section{Introduction}

COVID-19 is the latest evidence of epidemic disease capable of producing an extraordinarily large number of infections starting from a few \cite{nishiura2020closed}. According to Lippi and Plebani, \ac {covid19}, which originated in the city of Wuhan China on December 01, 2019, is a respirational and zoonotic disease, caused by a virus of the coronaviridae family~\cite{Lippi2020}. 
The Virus strain is severe acute respiratory syndrome coronavirus 2 (SARS-CoV-2), resulting in fever, coughing, breathing difficulties, fatigue, and myalgia. It may transform into pneumonia of high intensity. 

Towards successful diagnostic and cure of COVID-19, scientists in the field of molecular biology \cite{callaway2020} are working hard to find answers about its spreading and infecting by examining virus samples. Although, the disease strain is known, but, the vaccine is no where near. And it is essential to ensure strict mitigation actions so that virus can be contained. Already, most of the countries are taking different kinds of precautionary measures to cope with it so that the losses can be reduced.

From the experiences of China and South Korea, the countries able to contain the spread of the disease so far, it is learnt that \textit{social distancing} and \textit{testing} are the key factors. Another factor responsible of spreading of the disease worldwide was regional and global travellers. Although, air travel is almost suspended now, but, this initial shock and countries (and people) not taking it too seriously has taken many countries in Europe and North America to a real bad situation. 

In this overall scenario, after widespread suspension of air travels, Pakistan has taken various other measures to avoid spreading the infection. Schools and Colleges/Universities were closed from March 13, 2020, followed by other non-essential offices and services. The situation is been monitored on daily basis by the federal and provincial governments. The testing kits are being imported and more and more tests are conducted with each passing day. However, the number of tests is still quite low. By the time of this writing, most of the country is partially locked downed. However, it is hard to convince people to stay home and take precautionary measures. Many places are still covered with people.

Countries like Pakistan have four problems which make them more vulnerable than others; (i) a huge congested population, (ii) lack of medical facilities, (iii) poverty, and (iv) culture. Although, the spread of the disease is not that much as of today (March 31, 2020). But, the new cases are appearing continuously. And the next few week are very important.

To know what may happen, we need to model and simulate. Towards this, we model the dynamics of spread of the disease (the epidemic model), population data and mobility of people. We have used a system that is designed to to it.

The epidemic model presented in this paper is not novel. In fact, similar (or even same) models are already proposed in different fields of study. However, contextualization and implementation of the model in the current global and regional situation is significantly important. Already such studies are been taken up by the research groups working in this area \cite{chinazzi2020effect}. Through this paper, we have provided a focused analysis of the situation and asked important what-if questions, particularly in the context of Pakistan. However, the suggested method can be applied to any other country of the World, or even at the global level.

\section{COVID-19 Disease Spread Model}

\subsection{The Base Model}

The model is based on well-established state-transition systems that are being used to study epidemics for a long time. The simplest one are SI (susceptible-infectious) and SIR (susceptible-infectious-recovered) models \cite{brauer2012mathematical}. In both, an \textit{infectious} individual infects a \textit{susceptible} individual at a rate $\beta$. In SIR model, we also have a recovery rate ($\mu$) after which an infectious individual is \textit{recovered} permanently.  

It can easily been seen that simple SIR model does not fully grasp COVID-19. For epidemics like corona-virus, SIR model was extended to SEIR \cite{perez2009agent}, introducing a new state \textit{exposed} between susceptible and infectious. This state is also known as \textit{latent}, representing the period during which the individual has been infected but is not yet infectious himself. Therefore, we also have an exposed rate denoted by $\epsilon$. 

Still, the model needs further extension. The closet model representing the COVID-19 specifications is the one proposed for H1N1 epidemic \cite{balcan2009seasonal}. Like in H1N1, in COVID-19, we have two types of infectious individuals; one that show symptoms (Symp) and the other who do not show symptoms (ASymp). And, an exposed individual can transit to state \textit{infectious\_Symp} with rate $\epsilon$ or to a state \textit{infectious\_ASymp} with rate $1 - \epsilon$.   

\subsection{The COVID-19 Model}

\begin{sidewaysfigure} [!htbp]
	\centering
	 \includegraphics[width=0.99\textwidth]{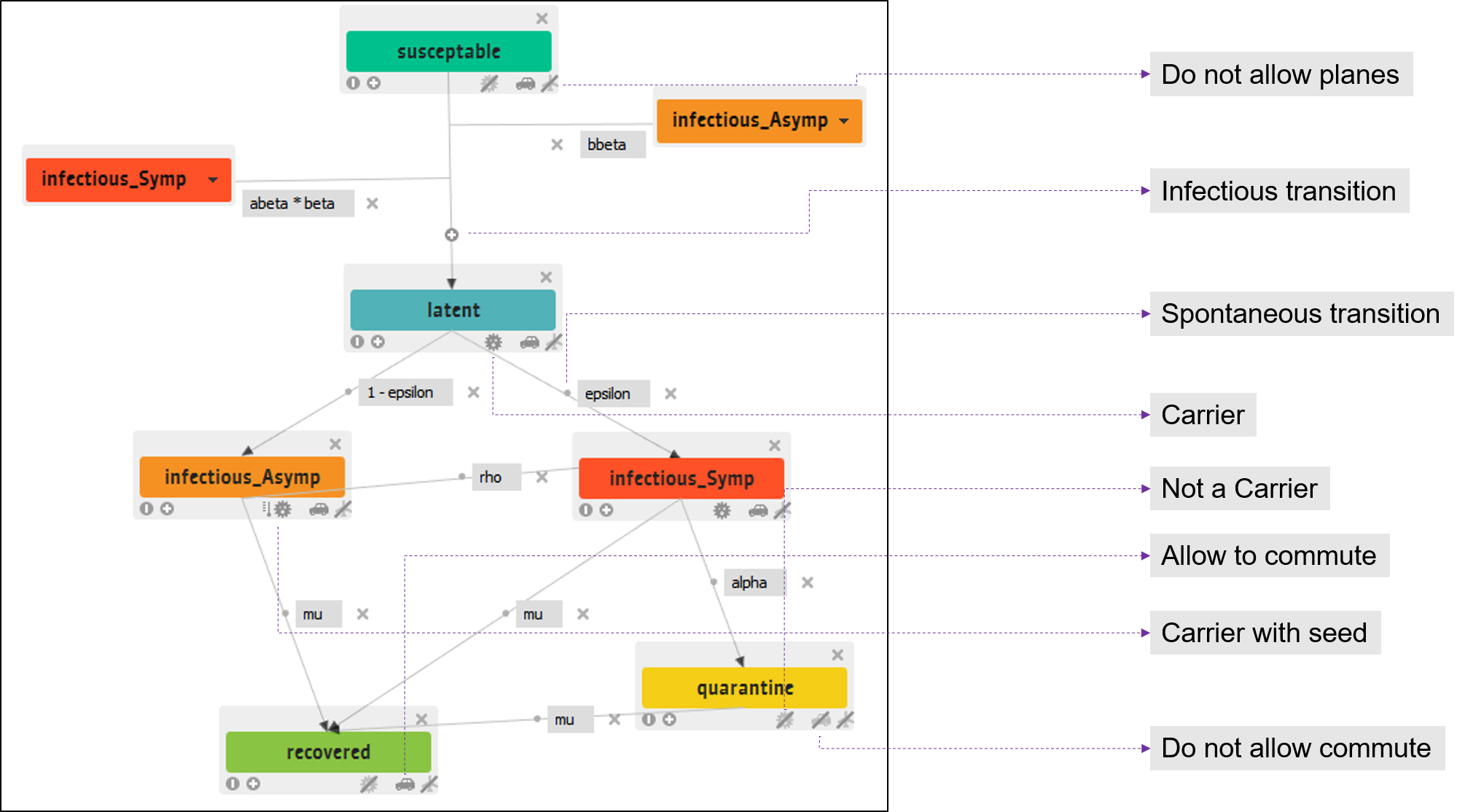}
\caption{The COVID-19 system dynamics model.}
\label{fig:model}
\end{sidewaysfigure}

The base model adapted from \cite{balcan2009seasonal} is further extended to incorporate a new state \textit{isolated} or quarantine. The isolated state represents the possibility of transferring an infectious individual with symptoms to isolation with a rate $\alpha$. The value of $\alpha$ is then used to represent preparedness of health system of a country (or globe). All three states infectious\_Symp, infectious\_ASymp, and isolated, transits to recovered state with same rate $\mu$. Also a transition from infectious\_ASymp to infectious\_Symp is made possible with a rate $\rho$. The final model is shown in Figure. \ref{fig:model}.

\section {Model Implementation}

\subsection{GLEAMviz Simulator}

The model is implemented in GLEAMviz \cite{van2011gleamviz}, the global epidemic and mobility model. GLEAMviz is a simulator which uses real-world data of population and mobility networks (both airways and commuting) on the server side. It integrates this data with the model developed by a user on the client side (similar to what we have presented in Figure. \ref{fig:model}). Hence, the simulation generated in data driven, in which a user is responsible of describing system dynamics model of the epidemic, whereas, all the relevant population and mobility global data in integrated at the server side. As a result, the time-series data of spread of epidemics is generated by the server system. 

In GLEAMviz client, the model is developed by showing transitions between different compartments (states). The model, conceptualized in the previous section, built on compartments and transitions is shown in Figure \ref{fig:model}. One aspect not explained yet is that there are two type of transitions. The \textit{infectious} transition is represented by ``+" sign, depicting an addition of infectious cases. There is only one instance -- from susceptible to exposed/latent -- like that. The other transitions are \textit{spontaneous} (represented by dot sign). The initial seed to the model is provided at compartment infectious\_Asymp, that is, few initial effectees as starting seed that do not possess any symptoms. 

\subsection {Model Specifications}

The are two functional modes of the model.

\subsubsection {Carrier-ship and Mobility}

An individual who is exposed to the virus (whether with symptoms or without) is the carrier of the virus. Therefore, a susceptible and recovered individual is not a carrier, while others are. Although, technically, an isolated/quarantined individual is a carrier, but, we have assumed that she/he is quarantined and is no longer able to transmit. Next are mobility possibilities. We have mimicked very recent situations to restrict or allow mobility. For example, considering all air traffic suspended, no compartment allows air travel. Whereas, all individuals who are not quarantined are allowed to commute locally. The commuting restrictions are further modified by using different transition variables.

\subsubsection{Transition Rates}

The following are the transition rates from one compartment to another. Note, that there are quite a few refinement in the base model. Also, for a Greek alphabet, it's symbol and text is used interchangeably. All rates vary between 0 and 1, inclusive.

\begin{itemize}
    \item \textbf{beta}: infectious rate that transforms susceptible to exposed. This happens under the influence of both individuals with or without symptoms. To differentiate, we have taken \textbf{bbeta} as the $\beta$ value for individual with no symptoms, and $\textbf{abeta} \times \beta$ as the value for individual with symptoms. In this way, we are able to relate the infections incurred in different situations.
    
    \item \textbf{epsilon}: rate of transiting from exposed to infectious state. An exposed individual can transit to state infectious\_Symp with rate $\epsilon$ or to a state infectious\_ASymp with rate $1 - \epsilon$. The value of $\epsilon$ is reciprocal of exposed period, which is equal to 5.2 days in our model \cite{diexpected}.  
    \item \textbf{rho}: rate of transiting from infectious\_ASymp to infectious\_Symp state. The value of $\rho$ is reciprocal of symptoms appearing period, which is equal to 2.3 days in our model \cite{diexpected}.  
    
    \item \textbf{mu}: rate of transiting from being infectious or isolated to recovered. The value of $\mu$ is reciprocal of infectious period. It is taken 30 days in case of COVID-19.
    
    \item \textbf{alpha}: rate of transiting from infectious\_Symp to isolated state.
\end{itemize}

\section {Parameterization and Cases}

The variations in beta, abeta, bbeta, and alpha make up different cases corresponding to different situations. The other variables (epsilon, rho and mu) are kept constant. Variations are introduced systematically based on what is observed in the last one month and what are possible actions of the future. We have categorized different situations (cases) based on the outcomes, which are: extremely good (case 1), extremely bad (case 2), and intermediate (case 3).

\begin{figure} [htp]
	\centering
	 \includegraphics[width=0.99\textwidth]{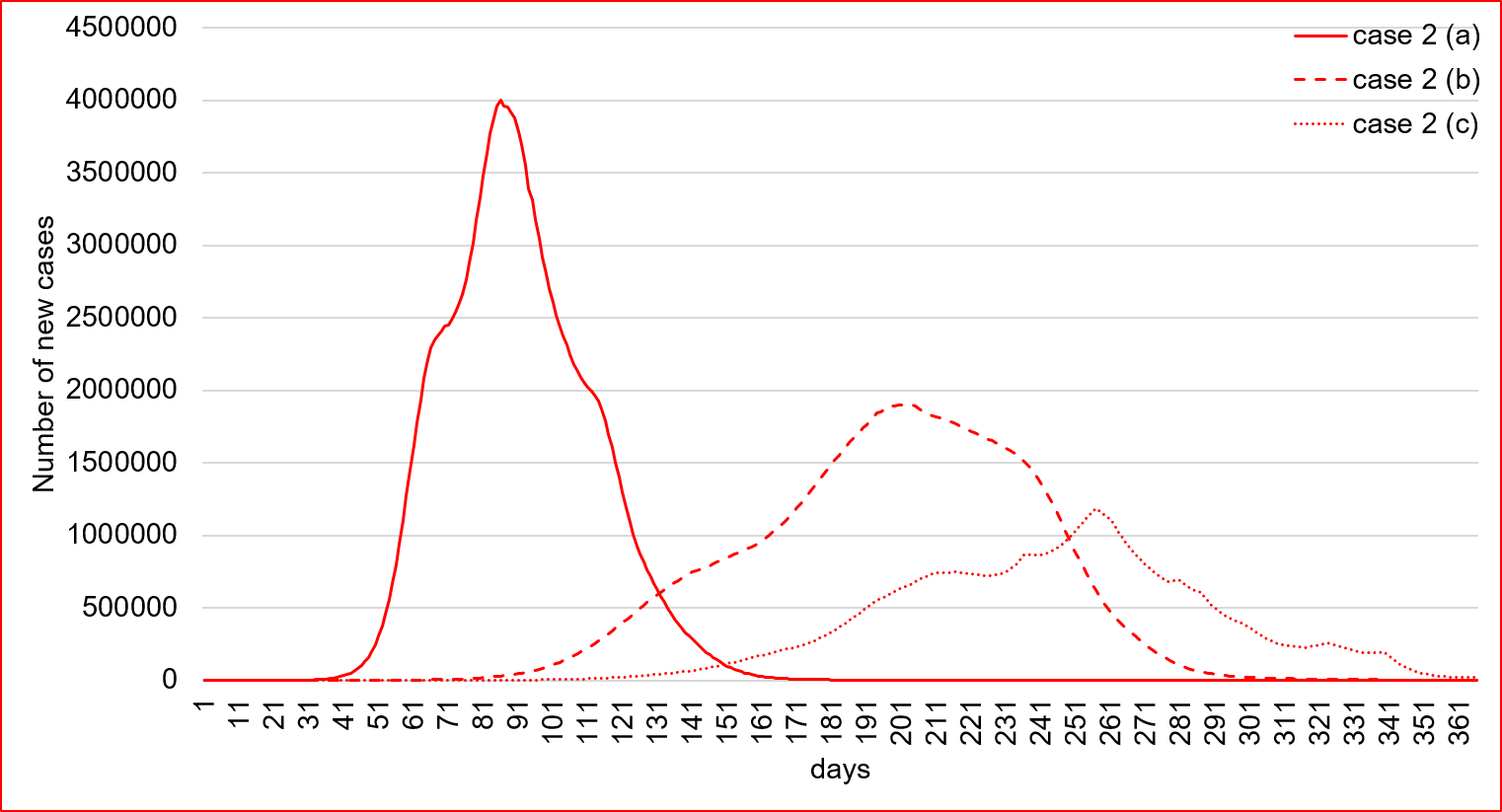}
\caption{Infections (with symptoms): case 2}
\label{fig:bad}
\end{figure}

\begin{figure} [htp]
	\centering
	 \includegraphics[width=0.99\textwidth]{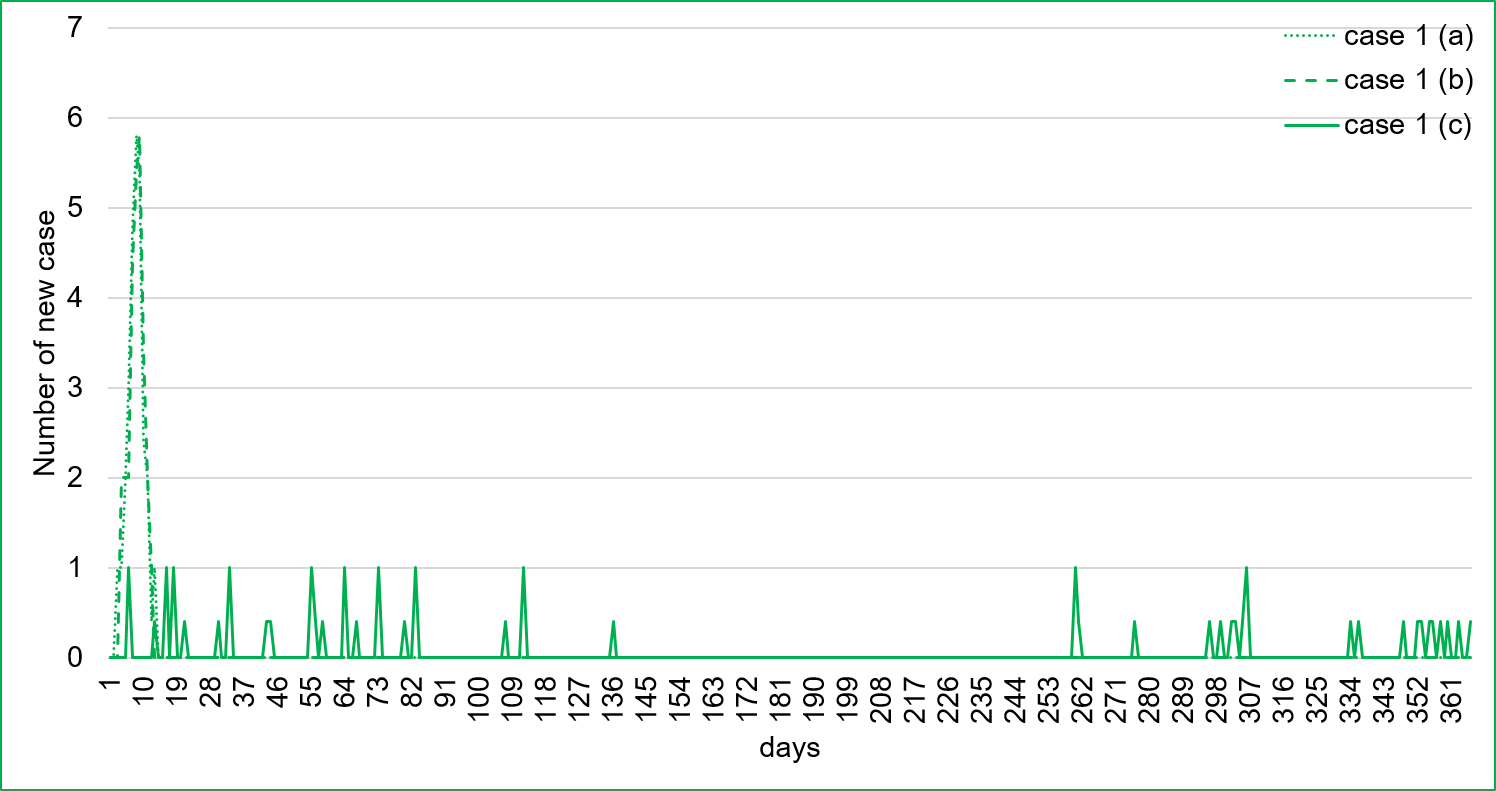}
\caption{Infections (with symptoms): case 1}
\label{fig:good}
\end{figure}

\subsection{Complete Inaction: Case 2 (a)} What can be worse than a complete inaction by the authorities? The default values assigned to the variables: $beta = bbeta = 0.5$, $abeta = 1$, and $alpha = 0.001$, were able to generate such a situation. In this case, $abeta \times beta = 1.0 \times 0.5$ (infectious rate incurred by infected individuals with symptoms) and $bbeta = 0.5$ (infectious rate incurred by infected individuals with no symptoms) both are 0.5, depicting the basic setting with no differentiation. The fact that the rate of getting isolated in really low ($alpha = 0.001$), depicts that there is no effort yet put by the authorities to contain the epidemic. In the context of Pakistan, we have put a few cases in Islamabad, Karachi and Gilgit as the starting cases which are not yet got any symptoms, and ran the simulation for a year, starting from February, 26, 2020 (when a few such cases were reported in the above mentioned cities). Even though the infection rate ($\beta$) is intermediate (only 1 out of 2 susceptibles are infected) and all flights are suspended (there are no outside influence), the results about spread of epidemic are really bad. 

\begin{figure} [htp]
	\centering
	 \includegraphics[width=0.99\textwidth]{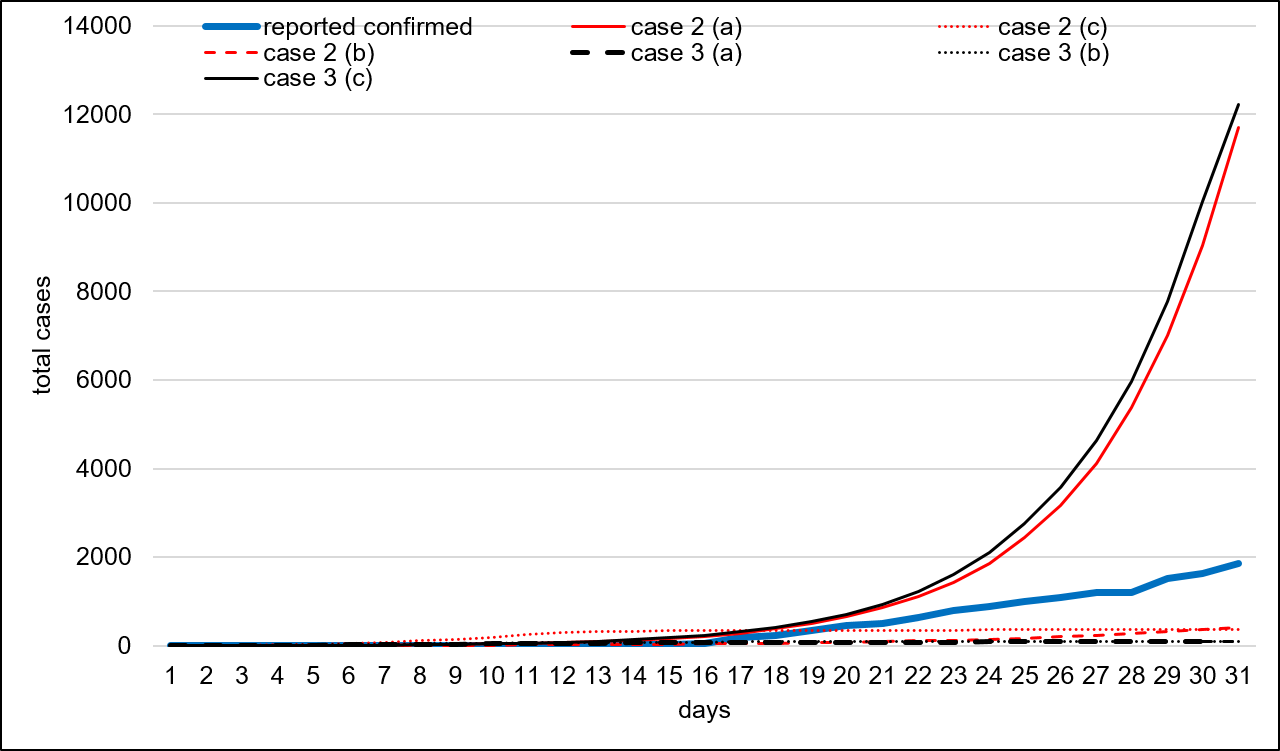}
\caption{Infections (with symptoms): comparison first 30 days}
\label{fig:comp30}
\end{figure}

\begin{figure} [htp]
	\centering
	 \includegraphics[width=0.99\textwidth]{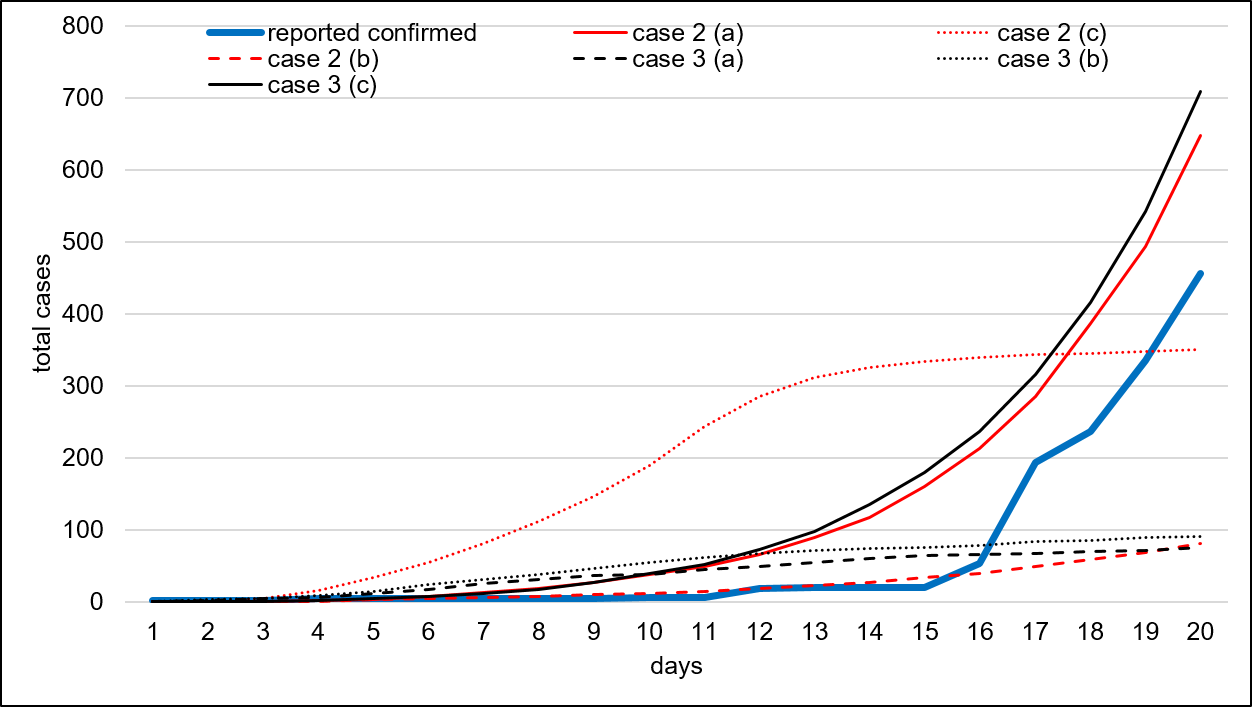}
\caption{Infections (with symptoms): comparison first 20 days}
\label{fig:comp20}
\end{figure}

This is case 2 (a) shown in  Figure \ref{fig:bad}. It suggests that the outbreak would be rapid and extreme, reaching 4 million cases per day after 75 days of the outbreak, and then it would start dropping. 80\% of the population would be affected. A comparison between real cases reported for the first 30 days and what the model generated (in case 2 (a)) is given in Figure. \cite{fig:comp30}. It is evident from Figure. \cite{fig:comp20} that for the first 20 days, the model generated comparable accumulated cases. But later the model's cases were quickly escalated, where the cases reported in real did not. The reason can be the lack of testing. As of today (March 31, 2020), we still have around 14,000 suspected cases (the cases without testing). Including these with 1850 confirmed cases, we see total number very close to the number predicted by the model. Nevertheless, we are not going to see this case as reality after the action been/being taken.

\subsection {Time-barred (short) extreme isolation and lock-down: Case 1 (a and b)} In \textbf{case 1 (a)}, we try to mimic a \textit{forced isolation} and a \textit{limited lock-down} (enforced social isolation). This happens one week after the identification of the first cases in Pakistan. Many countries tried to put such restrictions for two weeks only. We reproduced that by imposing if from March 3, 2020 to March 21, 2020. GLEAMviz provides an option to put exceptions in the form of rules to be applied for certain time period and for certain locations. The rules relate to setting the values based on mathematical expressions. For the above stated two weeks, we created exceptions, which are: abeta = 0.01, bbeta = 0.05, and alpha = 0.95.

What these values mean? The alpha = 0.95 means that infectious\_symp to isolated rate is 95\% leaving only 5\% patients to infect others. However, there is less probability of that happening due to lock down. Hence, we multiple abeta with beta to reduce it. bbeta is also reduced due to this reason. However, these reductions are quite strict and would be relaxed a bit in other cases.

These exceptions were only applied to cities of Pakistan where the cases were reported, assuming that so far not many small towns and places were locked down or have proper health infrastructure to identify infectious individuals or to isolate them. This seemed like a big problem. But, it turned out that it did not turned out to be that bad. The reason may be that, initially, a limited spread happened and limited isolation and lock-down in specific places was enough to contain the epidemic. 

The curve shown in Figure \ref{fig:good} of case 1 (a) simply shows that a few cases appeared in first week or so and then epidemic was eradicated. 

Particular to Pakistan's situation, thus, it can be concluded that if people coming from Taftan and from other countries, were properly isolated and locked down, the situation would have been entirely different from what we see now or expect in the future. However, unfortunately, like many other countries, Pakistan also could not reposed in this way. 

At the time of conducting our simulation, we were not sure about effectiveness of localized lock-down. Hence, we thought about full-scale lock-down and created \textbf{case 1 (b)}. We applied the above rules all across Pakistan. The results as shown against case 1 (b) in Figure \ref{fig:good} are not much different from case 1 (a).   

Nevertheless, we are not going to see these cases as reality after the initial inaction.

\subsection {Time-barred (longer) extreme isolation and localized lock-down with bulk cases: Case 2 (c)}

Moving more towards the reality, we changed case 1 (a) as following. After a few days of initial cases, we introduced bulk cases (patients travelling and entering into Pakistan) into most important cities of Pakistan. Initially many of these cases were gone undetected. Avoiding any presumed extreme values, we opted for almost what was reported about this at that time; a couple of dozens. These people were integrated with their families and many of them were infected. 

Many countries in the World are now aware that a lock-down of a few weeks would not be sufficient. Therefore, in this case, we also extended the lock-down to 45 days. Hence, the results of this case can be considered as nearest to reality according to current actions of the authorities. Unfortunately, the results of the simulation were not good. 

Remember, case 1 (a) is an idealized case, having following restrictions for 15 days: abeta = 0.01, bbeta = 0.05, and alpha = 0.95. In fact, we extended the lock-down time to 45 days. But, mere inductance of a dozen of patients into the population changed the tide. The results of case 2 (c) are shown in Figure \ref{fig:bad}. The problem with this graph is that the exponentiation of the cases start very late, almost after six months and that it reaches more than 1 million cases per day. The expected population that would be affected is almost half of case 1 (a) though; a total of 40\% of the population.  

Many countries are fearing about such a number that would be infected. Leaving the other 40\% of the population which would remain undetected, the number reaches to 80\% of the population again. This may suggest that whatever we do, the number cannot be reduced. However, restrictions give us enough time (compare case 2 (a) with case 2 (c)). The model does not predict deaths. But, if we have time, before a real exponential growth happens, we can better prepare ourselves to reduce fatality rate.

Aside from the discussion in the above paragraph, we proceeded to find other conditions that may give us hope. In the following, we present possible scenarios of the future, both natural and man-made.

\subsection {Optimistic Weather Intervention}

Although, we are not sure about it, but there is a high probability that the infections rate would drop when the weather warms up. This was achieved by reducing infection rates of the base case, case 2 (a). In the first case, beta and bbeta was set to 0.25 (half of the current infection rate) represented as \textbf{case 2 (b)}. In the second case,  beta and bbeta was set to 0.05 (almost none) represented as \textbf{case 1 (c)}. 

Unfortunately, case 2 (b) did not have much impact. This meant that infection rate of 0.25 is still high (see Figure \ref{fig:bad}). However, it has shifted the curve of case 2 (b) to the left when compared to case 2 (a). case 1 (c) produced a significant impact (see Figure \ref{fig:good}). However, without lock-down and isolation of patients, the epidemic would never stop and it would continue even after one year. This can be acceptable for some countries, but again, the authenticity of warm and humid weather disintegrating the virus remains to be proved. But, if it is proved and the impact is significant, the epidemic would die on its own. Nevertheless, we cannot plainly plan based on these assumptions.

\subsection {Realistic Interventions by the authorities}

\begin{figure} [htp]
	\centering
	 \includegraphics[width=0.99\textwidth]{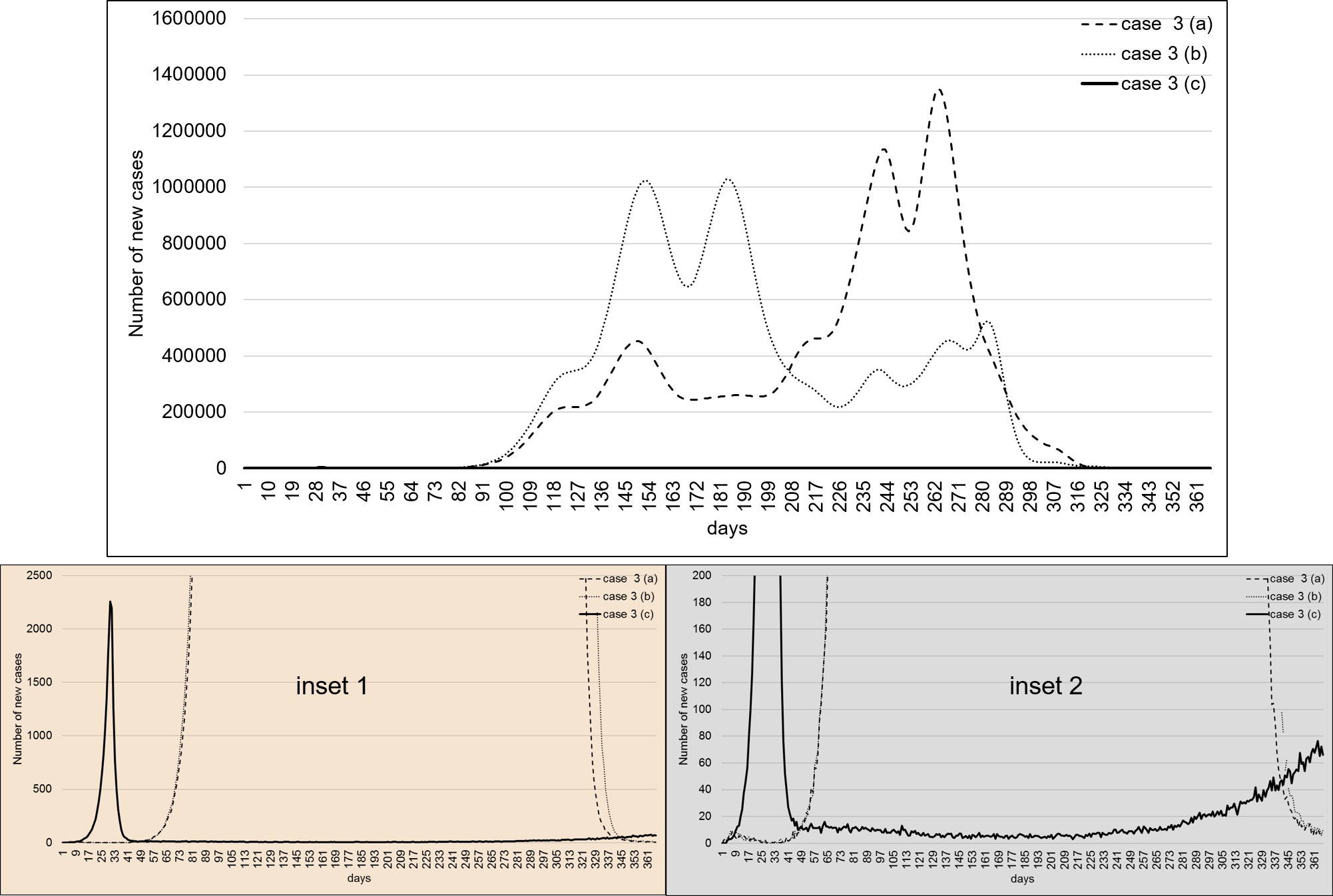}
\caption{Infections (with symptoms): case 3}
\label{fig:mid}
\end{figure}

According to the situation, the closest case we have seen so far is case 2 (c). The question is how can we do some actions which can reduce the devastation of case 2 (c). The problem with case 2 (c) is that it does not specify the proceedings of the last month as those happened in Pakistan (for example). It is too generalized. What happened in Pakistan is as follows:

\begin{enumerate}
    \item First few cases were reported without much action. However, the air travel was suspended right away.
    \item After one week, authorities took action and they started quarantine the people at selected places. This continued for 3 weeks. 
    \item Last week, there was a call of lock-down, but as of today, it is not that effective till now.
\end{enumerate}

Mapping the above situation, we created three variation of case 3. First, We ran the simulation for first week without any exceptions. Then we imposed a restriction of isolation (alpha = 0.95), first three weeks at selected places, and then rest of time all across the country. This also reduced the abeta value to 0.05. Similarly we imposed a lock-down following the same timeline. The lock-down was realized by taking bbeta value to 0.05 but only after one month and all across the country. The restrictions were applied for 45 days only and from today on wards. We named this case as \textbf{case 3 (a)}. This case represents real last month happenings with a hope that a restriction of 45 days would be enough. 

but, the results were not good as well as not bad. As shown in Figure \ref{fig:mid}, the outcome is very similar to case 2 (c) in terms of maximum new case per day and overall tally of patients. As we mentioned before, case 2 (c) was closet to the real situation. So is case 3 (a), as given in the above three points. But the results show that the number is overwhelmingly high. The only good thing is that it provides us enough time for preparation.  

What we have seen so far is that the restrictions are not 100\% effective. But, it does not mean that we may relax them. To demonstrate it, in \textbf{case 3 (b)}, we changed values of abeta and bbeta to 0.1. They never become 0.05. Results are really bad as shown in Figure \ref{fig:mid}. Overall, these two cases are very similar to case 2 (c). This also means that restricted lock-down and testing would not work.

The question is what can we do now, that is, after one month of epidemic break. Well, we should go for a complete lock-down and provide health services and quarantine facilities all across the country. We implemented this \textit{required} futuristic situation as follows. 

Being optimistic, we hope a complete lock down in coming weeks (that is from end of March) for a longer period than 45 days all across the country. Thus the value of bbeta would remain 0.05 from today on wards. But, if we will be able to provide sustained health services all across the country by identifying and isolating patients at villages level, the value of abeta can also be 0.05 for the whole year. Before today, for both variables, the values given are 0.1 (proving a benifit of doubt). We name this case as \textbf{case 3 (c)}. By looking at the graph of case 3 (c) in Figure \ref{fig:mid}, it is evident that at the start (almost as of today), the maximum has already reached at around 2200 cases a day (see inset 1 of Figure \ref{fig:mid}). After following the curve downwards, it almost stabilizes to 20 to 40 new cases per day later on (see inset 2 of Figure \ref{fig:mid}). This is quite manageable. But, it continues for a longer period, probably, beyond a year. 

The graphs is Figures \ref{fig:comp30} and \ref{fig:comp20} verify the two-phase response. In the first month, we could not do what was required, with numbers going up briskly. However, if we are able to tighten the restrictions even now, the coming weeks and month can turn out to be manageable. Still, there is a need of continuous restrictions, hoping that the cases per day would drop significantly low due to weather intervention.

\section {Discussion and Conclusion}

With a system dynamic model of epidemic spread, incorporated with population and mobility data, we performed simulation of many different cases of COVID-19 impact, representing different real situations. The data used was only about Pakistan, but following the method adopted, the simulation can be performed for any country or region. 

The exponential spread of COVID-19 virus has made all countries across the globe take preventive actions. Hence the first case (case 2 (a)) of unbounded spread, affecting almost 70\% of population was ignored. The epidemic is becoming more and more dangerous, partially due to absence of strict actions early on. The actions required are social distancing and healthcare management (including testing and isolation of patients). Many countries (including Pakistan) were not able to realize the impact of the situation and actually missed the train. Hence all the cases in category 1 (resulting in a real low number of people infected) were ignored as they are no more applicable.

Then, we encountered the real scenarios, specifically in the context of Pakistan. The first was case 2 (c). In this case, we mimicked 95\% isolation of infected patients with symptoms. The possibility of other 5\% infecting others was really less (only 2.5\%). The patients without symptoms was could infect others with a probability of 5\% only due to a lock down as well. Instead of turning out to be similar to the cases in category 1, the results showed that 40\% of population got infected, just due to inclusion of undetected bulk cases in the population. But the spread was delayed giving some time to prepare for it. Still peak of 1 million cases a day is a real nightmare. 

Case 3 (a) exactly matched what has happened in the last month after the breakout of the epidemic. Like case 2 (c), lock down was elevated after 45 days in case 3 (a) as well. Hence, case 3 (a) was not much different from case 2 (c). 

Lastly, we went for an indefinite lock down. Even, if the number of cases reached to 2000 a day (today on March 31, 2020) -- which can be true just due to overwhelmingly large number of suspected cases -- we saw a continuous decline after today. However, the restriction could not be lifted. As seen in other cases, when the restrictions were lifted after 45 days, the rate of spread of disease become exponential again.  

\bibliographystyle{unsrt}
\bibliography{main}

\end{document}